\documentclass[aps,pra,twocolumn,superscriptaddress,showpacs]{revtex4-1}
\usepackage[utf8x]{inputenc}

\usepackage[T1]{fontenc}

\usepackage{graphicx}
\usepackage{mathtools}
\usepackage{siunitx}
\usepackage{xcolor}
\usepackage{multirow}
\usepackage{hyperref}

\def\bra#1{\mathinner{\langle{#1}|}} 
\def\ket#1{\mathinner{|{#1}\rangle}} 
\def\braket#1{\mathinner{\langle{#1}\rangle}}

{\catcode`\|=\active 
  \gdef\Braket#1{\left<\mathcode`\|"8000\let|\bravert {#1}\right>}} 
\def\bravert{\egroup\,\vrule\,\bgroup}

\newcommand{\abs}[1]{\ensuremath{\vert {#1} \vert}}

\newcommand{\an}[1]{\ensuremath{\langle {#1} \rangle}}

\newcommand{\Sq}[1]{\ensuremath{\left[ {#1} \right]}}

\newcommand{\Prw}{\operatorname{Pr}^\text{w}}
\renewcommand{\Re}{\operatorname{Re}}
\newcommand{\W}[3]{\ensuremath{\prescript{}{#1}{\langle {#2} \rangle}_{#3}^\text{w}}}
\newcommand{\phm}{\phantom{-}}

\begin{document}

\title[Weak values to determine ``negative probabilities'' in a two-photon state]
{Using weak values to experimentally determine ``negative probabilities'' in a two-photon state with Bell correlations}
\author{B.~L. Higgins}
\affiliation{Centre for Quantum Dynamics and Centre for Quantum Computation and Communication Technology, Griffith University, Brisbane 4111, Australia}
\affiliation{Institute for Quantum Computing and Department of Physics \& Astronomy, University of Waterloo, Waterloo, ON N2L 3G1, Canada}
\author{M.~S. Palsson}
\affiliation{Centre for Quantum Dynamics and Centre for Quantum Computation and Communication Technology, Griffith University, Brisbane 4111, Australia}
\author{G.~Y. Xiang}
\affiliation{Centre for Quantum Dynamics and Centre for Quantum Computation and Communication Technology, Griffith University, Brisbane 4111, Australia}
\affiliation{Key Laboratory of Quantum Information, University of Science and Technology of China, Hefei 230026, China}
\author{H.~M. Wiseman}
\affiliation{Centre for Quantum Dynamics and Centre for Quantum Computation and Communication Technology, Griffith University, Brisbane 4111, Australia}
\author{G.~J. Pryde}
\email{G.Pryde@griffith.edu.au}
\affiliation{Centre for Quantum Dynamics and Centre for Quantum Computation and Communication Technology, Griffith University, Brisbane 4111, Australia}

\begin{abstract}

Bipartite quantum entangled systems can exhibit measurement correlations that violate Bell inequalities, revealing the profoundly counter-intuitive nature of the physical universe. These correlations reflect the impossibility of constructing a joint probability distribution for all values of all the different properties observed in Bell inequality tests. Physically, the impossibility of measuring such a distribution experimentally, as a set of relative frequencies, is due to the quantum back-action of projective measurements. Weakly coupling to a quantum probe, however, produces minimal back-action, and so enables a weak measurement of the projector of one observable, followed by a projective measurement of a non-commuting observable. By this technique it is possible to empirically measure weak-valued probabilities for all of the values of the observables relevant to a Bell test. The marginals of this joint distribution, which we experimentally determine, reproduces all of the observable quantum statistics including a violation of the Bell inequality, which we independently measure. This is possible because our distribution, like the weak values for projectors on which it is built, is not constrained to the interval $[0, 1]$. It was first pointed out by Feynman that, for explaining singlet-state correlations within ``a [local] hidden variable view of nature ... everything works fine if we permit negative probabilities''. However, there are infinitely many such theories. Our method, involving ``weak-valued probabilities'', singles out a unique set of probabilities, and moreover does so empirically.

\end{abstract}

\maketitle

\section{Introduction}

Entanglement is the most strikingly counter-intuitive property of quantum physics, and also underpins many quantum information technologies~\cite{Nielsen2000,Horodecki2009,Ladd2010,Zeilinger2010,Wiseman2010}. As first discussed by Einstein, Podolsky and Rosen~\cite{EPR35}, the quantum description of bipartite entangled systems 
is incompatible with the notion that systems have properties that exist locally, independent of measurement, and unaffected by distant events. This is formalized as the joint assumption of locality (that causal influences are lightspeed-limited) and realism (that systems possess objective properties  that determine measurement results). Following Bell's proof that this joint locality--realism assumption leads to empirically testable inequalities~\cite{Bell64}, many experiments (e.g.~\cite{Freedman72,Aspect82,Aspect82b,Weihs98,Kwiat99,Salart08,Scheidl10,Giustina2013}) have demonstrated that at least one of these assumptions must be false, contrary to our strong classical intuitions.

Tests of Bell inequalities typically involve measuring two pairs of observables, one pair on each system. Because the observables in each pair must be non-commuting (e.g. $\hat\sigma_X = \hat X$ and $\hat\sigma_Z = \hat Z$ of a quantum bit), measurement back-action makes it impossible to simultaneously measure all four combinations of observables. Rather, only one of the four combinations can be measured on each instance of the bipartite state. As particular measurement combinations are chosen, randomly and locally, throughout a Bell experiment, observed correlations in measurement results contradict the joint locality--realism assumption, evident as a violation of the Bell inequality under the conditions of standard probability theory~\cite{Fine1982}.

The assumption of locality and realism is equivalent to the existence of a joint probability distribution over all outcomes 
for all observables, even if these cannot be measured as relative frequencies because of measurement back-action. 
Implicit in this formulation is that every probability in the distribution must be between 0 and 1, inclusively. As first noted by Feynman in the context of singlet correlations which violate local realism~\cite{Feynman91}, for a local hidden variables (LHV) theory, 
``everything works fine if we permit negative probabilities.'' That is, 
by relaxing this condition and allowing for \emph{anomalous} probabilities---which includes both negative probabilities and 
their necessary correlate (because of normalization), probabilities greater than unity---the discrepancy between intuitive concepts of local realism and the actuality of quantum experiments can be made to vanish.  Feynman gave one procedure 
for constructing a local hidden variables theory based on anomalous probabilities, defining a complete joint distribution of anomalous probabilities. However, there are infinitely many such distributions, all of which satisfy the remaining conditions of probability and the experimental model.

One formalism which has proven useful for providing unique answers where standard quantum mechanics 
provides no answer, or many possible answers, is that of weak values~\cite{AAV88}. 
These have been applied experimentally to areas as diverse as 
elucidating the 3-box paradox~\cite{Resch2004b}, 
measuring Bohmian-like trajectories~\cite{Kocsis11}, and 
testing measurement--disturbance and complementarity relations~\cite{Mir07,Steinberg2012,Weston2013,Kaneda2014}, 
providing new ways to understand counter-intuitive and apparently paradoxical behaviours found in quantum systems. In weak measurements, quantum back-action is made arbitrarily small by having a weak coupling between system and  measurement probe. Thus, it is possible to perform a weak measurement of an eigen-projector for one observable ($\hat Z$, for example) followed by a strong measurement of a complementary observable (e.g.\ $\hat X$) on the same particle. Although the results of weak measurements have a large variance due to the weak coupling, a good signal to noise ratio can be achieved by averaging over a sufficiently large ensemble. Importantly, every member of the ensemble is subject to the same measurement.

Using this technique of a weak measurement of one projector, followed by a strong measurement of a different observable, one can ``observe the unobservable'', namely joint probabilities for the outcomes of incompatible measurements on an individual system. To be more precise, weak values provide \emph{empirically grounded}  ``weak-valued'' joint probabilities, which may of course be anomalous. In this paper,  we apply this idea to  polarization-entangled photons, to determine a \emph{unique} distribution of weak-valued joint probabilities for the complementary observables used in a Bell-type experiment, out of the infinite set of such distributions. Such anomalous weak-valued probabilities are encapsulated in a particular set of marginals of these distributions, and it is these marginals which we demonstrate experimentally. We stress that our experiment does not constitute a Bell test, but rather gives empirical meaning to Feynman's style of ``explanation'' for the paradoxical correlations in such a test. However, we do independently perform a Bell test on the same two-photon entangled states. (We do not attempt to close any of the applicable loopholes, however, as that is not the focus of this work.) This work was inspired by Hoffman's proposal~\cite{Hofmann2010,Hofmann2009} for inferring properties of a quantum system between preparation and the final strong measurements. Our weak-measurement technique is similar to that in other recent experiments~\cite{Kocsis11,Lundeen11,Dressel2011,Steinberg2012}.

\section{Applying weak measurement formalism to the CHSH observables}

To investigate quantum correlations 
through weak measurements we study a version
of Bell's inequality often used for experimental demonstrations, that due to 
Clauser, Horne, Shimony, and Holt (CHSH)~\cite{CHSH69}.
The CHSH inequality can be written as a bound on the expectation value of the CHSH parameter, which is 
the following correlation between measurement results on a pair of observables on each of two systems:
\begin{equation}
S_{\text{CHSH}} = (X + Z) P + (X - Z) Q. \label{eq:CHSHparam}
\end{equation}
Here $X, Z \in \{\pm1\}$ are random variables representing the results of a pair of measurements on Alice's system $A$, and $P, Q \in \{\pm1\}$ similarly for Bob's system $B$. In any individual experimental run, $S_\text{CHSH}$ may then take one of two values: $+2$ or $-2$. Assuming locality and realism, we can evaluate the expectation of this by summing over the 16 possible combinations of measurement outcomes, denoted by the corresponding lower-case letters:
\begin{equation}
\an{S_\text{CHSH}} = \sum_{x,z,p,q} [(x + z) p + (x - z) q] \Pr[x,z,p,q]. \label{eq:CHSHexpect}
\end{equation}
Note only one of $(x + z)p$ or $(x - z)q$ contributes to each term of the sum because one must be zero. If the probability of each outcome is $0 \leq \Pr[x,z,p,q] \leq 1$, it is simple to arrive at the CHSH inequality, $\abs{\an{S_\text{CHSH}}} \leq 2$.  Under strict experimental conditions, demonstration of a statistically certain violation of the CHSH inequality implies that the assumption of local realism must fail.

\begin{figure}
 \centering
 \includegraphics{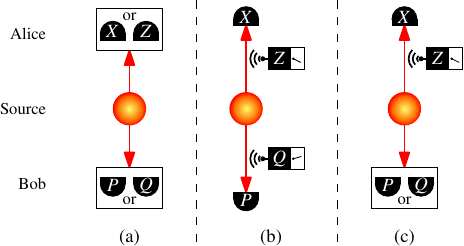}
 \caption{Measurements on entangled states. A source produces a pair of systems and sends one each to Alice and Bob, who each measure from two possible observables. In a Bell test, (a) Alice and Bob each choose one or the other observable to measure strongly for each system. Using weak measurements (b) no such choice is necessary, as Alice and Bob may collect statistics of every observable with a fixed configuration. Doing so, Alice and Bob may infer probabilities for simultaneous outcomes that cannot be measured directly. Alternatively (c), Alice may perform weak measurements while Bob performs strong measurements. Configurations (b) and (c) generate identical weak-valued joint probabilities for values of the CHSH parameter $S_\text{CHSH}$ owing to its factorization (see text).
}
 \label{fig:figure1}
\end{figure}

Quantum mechanics permits violation of this inequality when, e.g., Alice and Bob each possess one part of the maximally entangled 2-qubit singlet state, $\ket{\Psi^-} = (\ket 0_A \otimes \ket 1_B - \ket 1_A \otimes \ket 0_B)/\sqrt 2$. The CHSH parameter becomes the operator
\begin{equation}
\hat{S}_\text{CHSH} = (\hat{X} + \hat{Z}) \otimes \hat{P} + (\hat{X} - \hat{Z}) \otimes \hat{Q}. \label{eq:CHSHop}
\end{equation}
To maximize violation of the CHSH inequality~\cite{Cirel80}, Alice can use the Pauli observables $\hat X \equiv \ket 0 \bra 1 + \ket 1 \bra 0$, $\hat Z \equiv \ket 0 \bra 0 - \ket 1 \bra 1$, and Bob the observables $\hat P \equiv -(\hat Z + \hat X)/\sqrt 2$ and $\hat Q \equiv (\hat Z - \hat X)/\sqrt 2$ (figure~\ref{fig:figure1}). A simple calculation yields $\abs{\an{\hat S_\text{CHSH}}} = \abs{\bra{\Psi^-} \hat S_\text{CHSH} \ket{\Psi^-}} = 2\sqrt 2$, violating the CHSH inequality by a factor of $\sqrt 2$.

Feynman~\cite{Feynman91}, famous for ``thinking outside the box'', suggested a different way to interpret this result. 
He noted that it is possible to  
obtain the quantum probability distribution for measurements on a singlet state as follows. Let $V_\pm$ be the event that Alice obtains the result $\pm 1$ when measuring Pauli observable $\hat{V}$ and likewise $U_\pm$ for Bob \cite{Footnote1}.
Then the complete set of singlet correlations can be obtained by considering $\Pr[U_+,V_+]$ for all $U$ 
and $V$. Feynman shows that this can be written as 
\begin{equation}
\Pr[U_+,V_+] = \sum_{a,b}  P_{ab}\, \wp_a(V_+)\, \wp_b(U_+), 
\end{equation}
where $a$ and $b$ are local hidden variables, and $\wp_a(V_+)$ is the probability of event $V_+$ conditioned on the value of 
Alice's LHV $a$, and $\wp_b(U_+)$ similarly. Of course this equation could not possibly reproduce the singlet correlations, 
which disprove the possibility of any LHV theory, if it were not for the fact that Feynman allows the probability 
distribution over the hidden variables, $P_{ab}$, to take anomalous (e.g.\ negative) values. 
Reverting to the notation we used above, 
Feynman's model defines a joint distribution for observables $X, Z, P$, and $Q$, 
\begin{equation}
\Pr[x,z,p,q] = \sum_{a,b}  P_{ab} \wp_a(X_x)\,\wp_a(Z_z)\,\wp_b(P_p)\,\wp_b(Q_q).
\end{equation}
where $x$ (for example) takes values $\pm 1$ as above, and corresponds to the events $X_\pm$. This joint distribution 
allows violation of the CHSH inequality because it contains anomalous probabilities emanating from those in $P_{ab}$. 

Feynman's prescription is certainly not the only one for constructing an anomalous joint distribution yielding the CHSH correlations. In fact there are an infinite number of distributions $\Pr[x,z,p,q]$ consistent with these correlations. This is because it takes (imposing normalization) 15 real numbers to specify all the $\Pr[x,z,p,q]$, whereas in any experiment involving projective measurements, the full set of joint probabilities (such as $\Pr[x,p]$, etc.)\ yields only 8 independent real numbers, because of the no-signalling property of quantum mechanics~\cite{Ghirardi80}. (Each of the four joint probability distributions, like $\Pr[x,p]$, defines four probabilities, but one of each has one redundant probability from normalization, giving 12 parameters. No-signalling then gives four constraints, each one like $\sum_p \Pr[x=+1,p] = \sum_q \Pr[x=+1,q]$, leaving 8 free parameters. Note that these constraints are satisfied automatically by any normalized joint distribution $\Pr[x,z,p,q]$.)

In this paper, we describe a strategy that provides \emph{unique} values for the joint probabilities in (\ref{eq:CHSHexpect}), grounded in observation using weak measurements as the enabling tool.
The key feature of weak measurements is that they minimize measurement back-action. This allows measurement results to be obtained for all four variables ($X$, $Z$, $P$, and $Q$) for the same two-qubit system simultaneously. For example, weak measurements can be performed in the $\hat Z$ basis on subsystem $A$ and in the $\hat Q$ basis on subsystem $B$, before strongly measuring the subsystems in the $\hat X$ and $\hat P$ bases respectively. Through this process one can determine $\abs{\an{\hat S_\text{CHSH}}}$ using fixed measurements on every system in the ensemble. (This theoretical possibility was also noted in~\cite{Dressel2011}.)

Weak measurement results have large variance due to the weak coupling between the system and the probe. However, the formalism of weak values allows us to extract an empirically defined average value for the observables weakly measured on the ensemble. Within the overall ensemble of identically prepared states, subensembles of systems can be defined by postselection, i.e., by the outcome of the final strong measurement. For each such subensemble, the average value of the results of the weak intermediate measurement, of some observable $\hat{O}$ in general, is known as the \emph{weak value}~\cite{AAV88} of $\hat{O}$ over that subensemble, which  
observationally grounds any discussion about the value of $\hat{O}$ between preparation and measurement.
For qubits, every Hermitian operator is proportional to the identity plus a projector; the former has a weak value of one, and the latter has a weak value that can be termed a \emph{weak-valued probability}, $\Prw$, for the system to have the property represented by that projector. That is, for qubits, every measurement of a weak value is equivalent to measuring a weak-valued probability~\cite{Lund2010}.

To apply this idea to the CHSH inequality, first consider the weak measurement in the $\hat Z$ basis of a single qubit in state $\ket \psi$, followed by a strong measurement of $\hat X$. The latter can be equivalently framed as a postselection on the system being found in the measurement eigenstate $\ket{\phi(x)}$ corresponding to the result $x$. For this postselected ensemble, the weak value of $\hat Z$ is the expectation value of the results of the weak measurements in the $\hat Z$ basis, in the limit that the measurement strength goes to zero. Using the notation of~\cite{Wiseman02},
the weak value of $\hat{Z}$ for this pre- and post-selection evaluates to~\cite{AAV88}
\begin{equation}
 \W{\phi(x)}{\hat{Z}}{\psi} \equiv \Re\Sq{\frac{\braket{\phi(x) | \hat Z | \psi}}{\braket{\phi(x) | \psi}}}.
\end{equation}
Weak values are unusual in that they can take values outside the spectrum of the measured operator (here $\hat Z$). 
This is possible because the values of individual weak measurements are not constrained to the eigenvalues of the measured operator. Weak values have proven 
useful in increasing measurement precision where the resolution of the measuring device is otherwise the limiting factor~\cite{Hosten08,Dixon2009,Brunner2010,Starling2010,Hofmann2011}, in resolving a number of quantum mechanical paradoxes (e.g.~\cite{Kocsis11,Wiseman02,Steinberg95,Resch04,Brunner04,Resch2004b,Pryde2005,Mir07,Lundeen09,Yokota2009}) and investigating macrorealism on one~\cite{Williams2008,Goggin2011} or two~\cite{Dressel2011} systems.

The weak value $\W{\phi(x)}{\hat{\Pi}_z}{\psi}$ of the $\hat Z$ basis projectors $\hat{\Pi}_z$ can be interpreted as a weak-valued probability of obtaining the outcome $z$, given the $\ket \psi$ input state, conditional on finally postselecting the $\ket{\phi(x)}$ state indicating an outcome $x$ in the $\hat X$ basis, i.e.\ $\Prw[z | x, \psi] = \W{\phi(x)}{\hat{\Pi}_z}{\psi}$. Because the weak measurement's back-action on the system is negligible, it follows that the weak-valued joint probability of obtaining both $X=x$ and $Z=z$ outcomes is
\begin{align}
 \Prw[x, z | \psi] &= \Prw[z | x, \psi] \Pr[x | \psi] \\\nonumber
 &= \W{\phi(x)}{\hat{\Pi}_z}{\psi} \Pr[\phi(x) | \psi], \label{eq:joint1qb}
\end{align}
where $\Pr[x | \psi] = \Pr[\phi(x) | \psi] = \abs{\braket{\phi(x) | \psi}}^2$. Weak measurements thereby allow us to ascertain pseudo-probabilities for outcomes that are not directly obtainable by strong measurements.

Similarly, the weak-valued joint probabilities for the two-qubit case are
\begin{equation}
\Prw[x, z, p, q | \psi] = \W{\phi(x)}{\hat{\Pi}_z\otimes\hat{\Pi}_q}{\psi} \Pr[\phi(x, p) | \psi], \label{eq:joint2qb}
\end{equation}
where $\hat{\Pi}_q$ represents the $\hat Q$ basis projectors (on Bob's qubit) corresponding to the outcome $q$, and the postselection state $\ket{\phi(x,p)}$ depends on strong measurement outcomes of both Alice's ($X=x$) and Bob's ($P=p$) qubits. Note that all of the quantities required to find the weak-valued joint probability could be determined by experimental correlation statistics of the joint measurement.

We can thus calculate the weak-valued joint probabilities of each of the 16 outcomes of measurements on the entangled state $\ket{\Psi^-}$. For measurements maximizing the CHSH inequality violation, they each take one of four possible values, $(2 + \sqrt 2)/16$, $\sqrt 2/16$, $(2 - \sqrt 2)/16$, or $-\sqrt 2/16$, while each of the 16 corresponding outcomes results in a value for the CHSH parameter, $S_\text{CHSH}$, of ${\pm} 2$ by (\ref{eq:CHSHparam}). Thus we find that the total weak-valued probability for obtaining the positive value of the CHSH parameter, $S_\text{CHSH} = +2$, is $(1 + \sqrt 2)/2 \approx 1.207$, while the weak-valued probability for the negative value, $S_\text{CHSH} = -2$, is $(1 - \sqrt 2)/2 \approx -0.207$. It is easy to see that the contribution to the expectation value (\ref{eq:CHSHexpect}) is therefore positive in both cases, giving $\abs{\an{S_\text{CHSH}}} = 2\sqrt 2$ as expected~\cite{Hofmann2009}. 
As it turns out, these predicted values are quite different from the solution happened upon by Feynman from considering his particular model of anomalous probabilities motivated by his LHV model.

These anomalous (i.e.\ outside the range $[0,1]$) weak-valued probabilities may at first seem nonsensical. Indeed, they cannot be measured as relative frequencies in the laboratory---instead they must be inferred from weak measurement data, as we do below. They arise simply from the fact that they stand for strong measurement results which cannot be physically obtained (events that cannot actually take place) because of measurement back-action. As shown previously~\cite{Steinberg2012, Lund2010, Mir07, Lundeen09}, such anomalous probabilities are nevertheless useful in probability accounting for physically realizable events.

\section{Experimental apparatus}

\begin{figure}
 \centering
 \includegraphics[width=\linewidth]{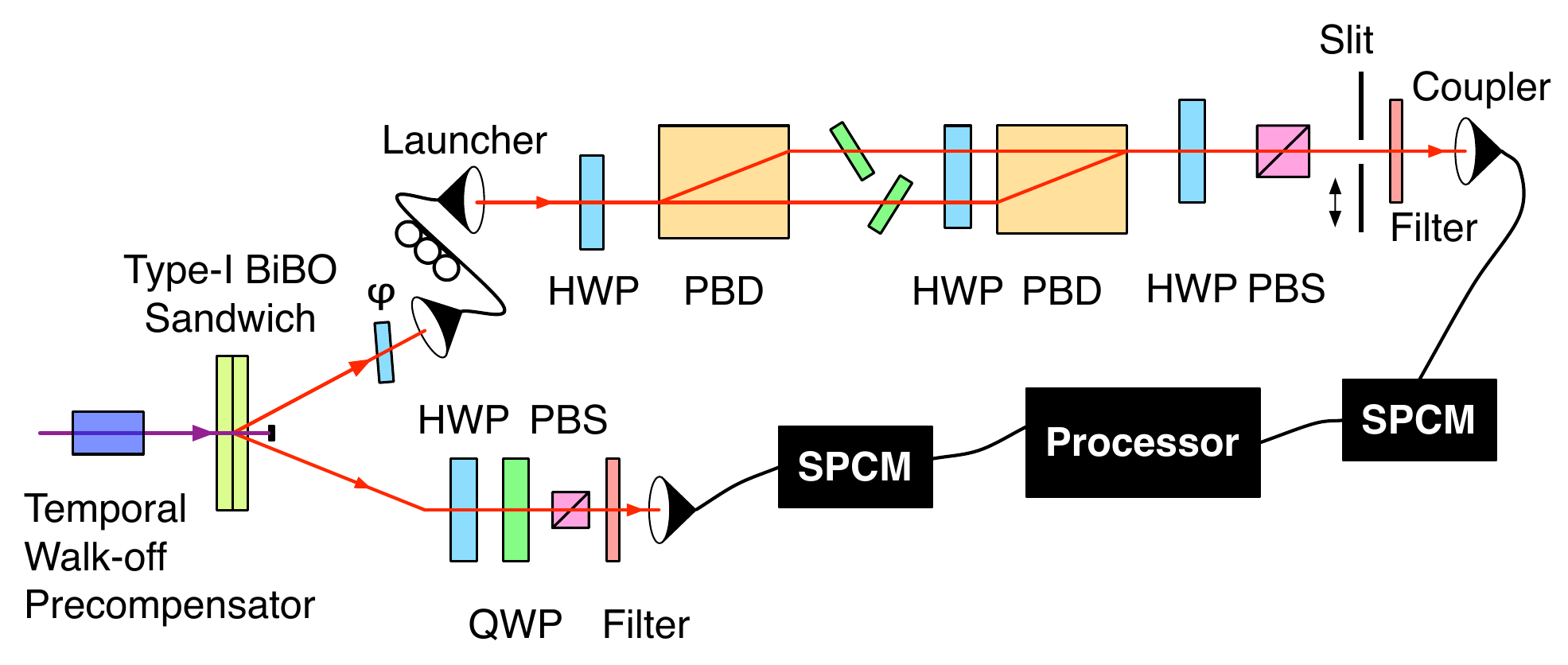}
\caption{Experimental apparatus. Pairs of entangled photons are generated by a type-I bismuth borate source. One of the photons undergoes strong measurement, the other is coupled to single-mode optic fibre and directed to the weak measurement apparatus, implemented as a polarization interferometer. A tilted optical plate is placed in each interferometric arm, causing a polarization-dependent transverse shift of the photon's spatial Gaussian mode, with polarization postselection following. The mode intensity is sampled by a scanning slit, and is collected into a multimode optical fibre (after passing through telescoping lenses, not shown), connected to a single-photon counting module (SPCM).} \label{fig:figure2}
\end{figure}

We now present our experimental demonstration of these anomalous weak-valued probabilities using photons as shown in figure~\ref{fig:figure2}. Firstly we note that if Alice could obtain both $X$ and $Z$ measurement results simultaneously, then by (\ref{eq:CHSHparam}) either $P$ or $Q$ of Bob's results would not matter for each shot. The only empirically relevant weak-valued probabilities, in terms of contribution to the CHSH parameter $S_\text{CHSH}$, are then either $\Prw[x, z, p]$ or $\Prw[x, z, q]$. We may therefore simplify the experiment by performing weak measurement of only one photon of each pair (Alice's), rather than both simultaneously (figure~\ref{fig:figure1}), with the weak-valued probabilities for each possible value of the CHSH parameter remaining unchanged. 

Pairs of polarization-entangled photons, having high fidelity with $\ket{\Psi^-}$, are generated via spontaneous parametric downconversion using a type-I `sandwiched pair' source~\cite{kwiat09}, pumped by \SI{410}{\nm} light from a mode-locked frequency-doubled Ti:sapphire laser. Two thin (\SI{0.6}{\mm}) bismuth borate (BiBO) crystals, one cut for downconversion of horizontally polarized incident light, the other cut for vertically polarized incident light, are placed back-to-back. Pumping by diagonally polarized light induces coherent downconversion from the two crystals resulting in the two-photon polarization state $(\ket H \ket H + e^{i\varphi} \ket V \ket V)/\sqrt 2$, for some constant phase $\varphi$. By changing the polarization of the pump light we can adjust the proportion satisfying the phase matching conditions of each crystal, thereby tuning the degree of entanglement generated. For a pump polarization angle $\theta$, the resulting state is $\ket{\psi(\theta)} = \sin\theta \ket H \ket H + e^{i\varphi} \cos\theta \ket V \ket V$ with a tangle (squared concurrence)~\cite{Munro2001} of $\sin^2 2\theta$.

Each photon of the generated pair is assigned to either Alice or Bob. Bob's photon immediately undergoes strong measurement, either $\hat{P}$ or $\hat{Q}$, by postselection using a quarter-wave plate (QWP), half-wave plate (HWP), and polarizing beamsplitter (PBS). Alice's photon passes through a HWP set to its optic axis, tilted around vertical such that it compensates for $\varphi$ and (in the condition of a diagonally polarized pump beam) achieves a state having high fidelity with $\ket{\Psi^-}$ at the measurement apparatus. The photon is guided to the weak measurement apparatus using a single-mode optical fibre, providing a Gaussian spatial mode. It passes through a HWP which determines the basis of the weak measurement---we initially set this wave plate to its optic axis, thereby implementing weak measurement in the $\hat{Z}$ basis.

Weak measurements are achieved by engineering a polarization-dependent displacement, $\Delta r$, significantly smaller than the width of the transverse Gaussian mode~\cite{Ritchie91}. To achieve this displacement, we construct a partially spatially-mismatched polarization interferometer. A polarizing beam displacer (PBD) separates the horizontal and vertical polarizations of the photon into two parallel spatial modes. An optical plate is placed in each mode and tilted approximately equally in opposing directions, causing small opposite displacements of the two modes following Snell's law. For our apparatus it was logistically convenient to use QWPs set at their optic axes. A HWP then flips horizontal and vertical polarizations and the modes are (partially) recombined by the final PBD. A large Gaussian beam width, relative to the size of the displacement, ensures that decoherence due to measurement back-action is negligible.

Alice's photon is then postselected in the conjugate basis using a HWP at $\pm\ang{22.5}$ from its optic axis followed by a PBS. A slit, seated on a motorized translation stage, is scanned in the direction transverse to the beam, allowing us to sample the photon flux of the Gaussian profile at any point $r$. The photon passes through telescoping lenses (not shown in figure~\ref{fig:figure2}) and a \SI{3}{\nm} full-width at half maximum (FWHM) interference filter, and is coupled into a multi-mode fibre guided to a single-photon counting module (SPCM). (Similarly, following postselection, Bob's photon also passes through a \SI{3}{\nm} interference filter, and is coupled into a single-mode optical fibre guided to a SPCM.) Alternatively to postselection, one could also perform strong two-outcome polarization measurements. In either case, integration of the photon flux (in coincidence with Bob's photons) allows us to determine the expectation of the weak polarization measurement, and thus the weak-valued joint probability of each polarization outcome of Alice's and Bob's photons. Assigning horizontal and vertical polarizations to the $Z=+1$ and $Z=-1$ states, respectively, we obtain weak measurements in the $\hat Z$ basis, with postselection in the $\hat X$ basis.

Detecting Alice's and Bob's photons within a coincidence window of ${\approx}\SI{3}{\ns}$ helps ensure high-fidelity entangled two-photon states, however some accidental coincident detections remain. We estimate these accidental count rates by simultaneously recording detection events of the two SPCMs coincident when displaced in time by the pump pulsing period. All such detections thus arise from uncorrelated events. We subtract these accidental coincident counts from our results.

We use a slit of width approximately \SI{350}{\um}, scanned over a \SI{3.5}{\mm} range in steps of \SI{87.5}{\um}, counting for \SI{10}{\s} at each step. This is performed for each of the eight total postselections for $X$, $P$, and $Q$, in turn. The process is repeated 70 times, ensuring that any drift in pump power is experienced equally (approximately) for each outcome. To calculate the joint probabilities it is necessary to characterize the centroid positions $r_H$ and $r_V$ associated with horizontally and vertically polarized photons exiting the interferometer, respectively. This is done by additionally postselecting horizontal and vertical photon polarization states, making a total of ten measurement conditions in the experiment. We fit a Gaussian function to the count rates of each characterization---the centres of these fits define our $r_H$ and $r_V$ estimates.

Because $r_V > r_H$ in our apparatus, it is convenient to write the displacement $\Delta r = r_V - r_H$. 
Measuring the position of the Gaussian profile for a given photon state corresponds to a $\hat{Z}$ measurement. In typical spatial-mode qubit encodings, the modes are sufficiently far apart that the Gaussian profiles are approximately orthogonal, and the measurement outcomes thus correspond to strong measurements. The expectation value for these measurements may be written
\begin{equation}
\an{\hat{Z}} = \int^{\infty}_{-\infty} \frac{r_{H}+r_{V}-2r}{\Delta r} \wp(r) \, dr, \label{eqn:expecZ}
\end{equation}
where $\wp(r)$ is the (Gaussian) probability density of detecting a photon at a position $r$. (For notational convenience, we omit the explicit dependence on the initial state $\ket \psi$.) As $\Delta r$ is reduced towards zero, the measurement becomes weak and the uncertainty of individual outcomes increases due to the overlapping Gaussian distributions for each outcome. In our apparatus, $\Delta r \approx \SI{150}{\um}$, in comparison to the ${\approx}\SI{820}{\um}$ FWHM of our Gaussian beam. Under these conditions, we calculate the measurement strength, quantified by the \emph{knowledge} $K$~\cite{Pryde2005}, to be about 0.024---sufficiently small to demonstrate our weak-valued probabilities.

In the absence of postselection, the expectation value $\an{\hat{Z}}$ is the same in both the weak and strong measurement cases. Therefore,
\begin{equation}
\sum_z z\Prw[z] = \int^{\infty}_{-\infty} \frac{r_{H}+r_{V}-2r}{\Delta r} \wp(r) \, dr. \label{eqn:WexpecZ}
\end{equation}
Using the fact that the two weak-valued probabilities in (\ref{eqn:WexpecZ}) must sum to one, it follows immediately that
\begin{equation}
 \Prw[Z=+1] = \frac{1}{\Delta r}\int^{\infty}_{-\infty}(r_V - r) \wp(r) \, dr, \label{eqn:PRZP}
\end{equation}
and
\begin{equation}
 \Prw[Z=-1] = \frac{1}{\Delta r}\int^{\infty}_{-\infty}(r - r_H) \wp(r) \, dr. \label{eqn:PRZM}
\end{equation}

The postselection outcomes become extra conditions on (\ref{eqn:PRZP}) and (\ref{eqn:PRZM}). Supposing Bob measures $\hat P$, Alice $\hat X$, then $\wp(r)$ becomes $\wp(r,x,p)$, and $\Prw[z]$ becomes $\Prw[z,x,p]$. The weak-valued joint probabilities can be calculated from experimental counting statistics by considering (\ref{eqn:PRZP}) and (\ref{eqn:PRZM}). For a finite slit width, the integral becomes a sum approximation over the range of measured positions as $dr$ becomes $\delta r$,
\begin{equation}
 \Prw[Z=+1,x,p] \approx \frac{1}{\Delta r} \sum_{r} (r_{V}-r) \wp(r,x,p) \delta r,
\end{equation}
and similarly for $\Prw[Z=-1,x,p]$.

The value of $\wp(r,x,p)$ cannot be measured directly, but must instead be estimated from photon detections. Let $C(r,x,p)$ represent the count rate of these detections. Then $\wp(r,x,p) \delta r \approx C(r,x,p) / C_\text{T}$, where $C_\text{T} = \sum_{r,x,p} C(r,x,p)$ is the total number of coincident photon detections over all the outcomes of the measurements in $\hat X$ and $\hat P$. The weak-valued joint probability can therefore be estimated by
\begin{equation}
 \Prw[Z=+1,x,p] \approx \frac{\sum_{r} (r_{V}-r) C(r,x,p)}{\Delta r \sum_{r,x,p} C(r,x,p)},
\end{equation}
and similarly for $\Prw[Z=-1,x,p]$, $\Prw[Z=+1,x,q]$, and $\Prw[Z=-1,x,q]$.

\section{Results}

\begin{table}[tb]
\begin{ruledtabular}
\begin{tabular}{@{}*{6}{c}}
\multicolumn{4}{c}{Outcome} & \multirow{2}{*}{$S_\text{CHSH}$} & \multirow{2}{*}{$\Prw$} \\ \cline{1-4}
$x$ & $z$ & $p$ & $q$ \\ \hline
$+1$ & $+1$ & $+1$ &  & $+2$ & $\phm0.332\pm0.005$ \\
$+1$ & $+1$ & $-1$ &  & $-2$ & $-0.030\pm0.001$ \\
$+1$ & $+1$ &  & $+1$ &  & $\phm0.149\pm0.004$ \\
$+1$ & $+1$ &  & $-1$ &  & $\phm0.152\pm0.002$ \\
$+1$ & $-1$ & $+1$ &  &  & $\phm0.099\pm0.005$ \\
$+1$ & $-1$ & $-1$ &  &  & $\phm0.103\pm0.002$ \\
$+1$ & $-1$ &  & $+1$ & $+2$ & $\phm0.244\pm0.004$ \\
$+1$ & $-1$ &  & $-1$ & $-2$ & $-0.048\pm0.002$ \\
$-1$ & $+1$ & $+1$ &  &  & $\phm0.131\pm0.002$ \\
$-1$ & $+1$ & $-1$ &  &  & $\phm0.148\pm0.004$ \\
$-1$ & $+1$ &  & $+1$ & $-2$ & $-0.042\pm0.002$ \\
$-1$ & $+1$ &  & $-1$ & $+2$ & $\phm0.320\pm0.004$ \\
$-1$ & $-1$ & $+1$ &  & $-2$ & $-0.051\pm0.002$ \\
$-1$ & $-1$ & $-1$ &  & $+2$ & $\phm0.276\pm0.005$ \\
$-1$ & $-1$ &  & $+1$ &  & $\phm0.131\pm0.002$ \\
$-1$ & $-1$ &  & $-1$ &  & $\phm0.087\pm0.004$ \\
\end{tabular}
\end{ruledtabular}
\caption{Measured weak-valued probabilities $\Prw$ for the 16 possible values of the observables $X$, $Y$, $P$ and $Q$, for the entangled state produced in the experiment. These are obtained from weak measurements in the $\hat Z$ basis. Not all conditions contribute to the CHSH parameter---appropriate summation of these values for the conditions satisfying $S_\text{CHSH} = +2$ and $S_\text{CHSH} = -2$ leads to positive and negative weights of $1.172\pm0.008$ and $-0.171\pm0.002$, respectively. Since these multiply terms of opposite sign, both give positive contributions to CHSH parameter $\an{S_\text{CHSH}}$, leading to violation of the CHSH inequality.} \label{tab:probs}
\end{table}

The photon source performance is quantified by conducting quantum state tomography~\cite{Kwiat04} with the slit removed and a QWP inserted immediately before the final HWP. With a two-photon state of measured Bell-state fidelity $0.948 \pm 0.002$ and concurrence $0.917 \pm 0.003$ (tangle $0.841 \pm 0.004$), we expect to be able to achieve a CHSH value of $|\langle S_{\text{CHSH}}\rangle|\approx2.67$. For this state, the experimentally determined weak-valued probabilities are given in table~\ref{tab:probs}. The weak-valued probabilities of the positive and negative values of the CHSH parameter are determined by taking the appropriate sums of these results. They are $1.172 \pm 0.008$ for $S_{\text{CHSH}} = +2$ and $-0.171 \pm 0.002$ for $S_{\text{CHSH}} = -2$, resulting in $\abs{\an{S_{\text{CHSH}}}} = 2.686 \pm 0.017$, violating the CHSH inequality by more than 40 standard deviations. (Note that we do not claim this demonstrates a violation of local realism, as the nature of our apparatus cannot support such a conclusion.)

\begin{figure}[tb]
 \centering
 \includegraphics{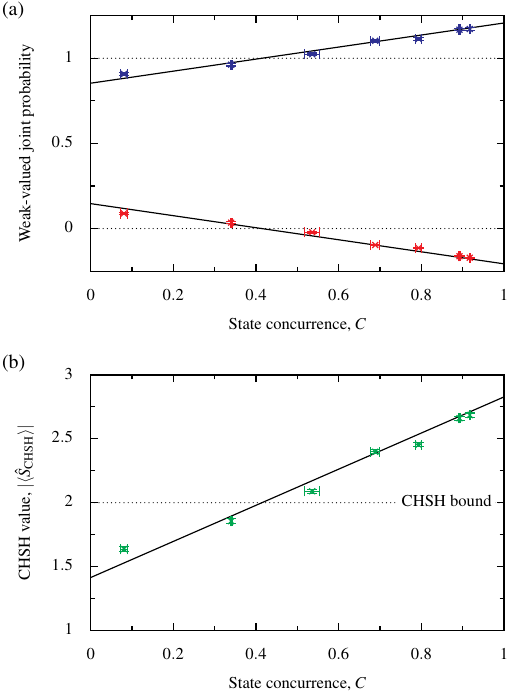}
 \caption{Experimental results for states of various degrees of entanglement, quantified by the concurrence $C$. (a) Weak-valued joint probabilities of outcomes corresponding to $S_\text{CHSH} = +2$ (upper curve) and $S_\text{CHSH} = -2$ (lower curve). (b) The CHSH value $|\langle\hat{S}_{\text{CHSH}}\rangle|$ calculated using the weak-valued joint probabilities. In both plots, the experimentally measured values (points) closely match the ideal theoretical values (lines) as the concurrence of the state varies. Horizontal error bars represent the range of concurrence values measured via tomography taken before and after data collection. Vertical error bars span one standard deviation, calculated from the (Poissonian) photon counting statistics.}
 \label{fig:figure3}
\end{figure}

Our choice of weakly measuring in the $\hat Z$ basis is arbitrary---we may instead measure $\hat X$ weakly before a strong $\hat Z$ measurement and obtain essentially identical expected weak-valued joint probabilities. To demonstrate this, we change the basis of the measurements by setting the initial HWP in the weak measurement device to \ang{22.5} from its optic axis. In this experimental condition we observe a concurrence of $0.926 \pm 0.003$ (tangle $0.857 \pm 0.005$, deviating slightly from previous values due to a drift of the photon source during the period between measurements), and weak-valued probabilities of $1.147 \pm 0.008$ for $S_{\text{CHSH}} = +2$ and $-0.140 \pm 0.002$ for $S_{\text{CHSH}} = -2$. This gives a CHSH value of $2.574 \pm 0.016$. Despite the improved entanglement for this state, sensitivity to imprecision in the manual setting of the change-of-basis HWP leads to slightly reduced magnitudes of the measured weak-valued probabilities.

With weak measurements in the $\hat Z$ basis, we also perform the experiment for states of various entanglement. Theoretically, for the states we produce with concurrence $C$, the CHSH value goes as $\abs{\an{\hat S_\text{CHSH}}} = (C + 1)\sqrt 2$, while the weak-valued probabilities of outcomes corresponding to positive and negative values of $S_\text{CHSH}$ go as $\Prw[S_\text{CHSH} = +2] = (2 + (C + 1)\sqrt{2})/4$ and $\Prw[S_\text{CHSH} = -2] = (2 - (C + 1)\sqrt{2})/4$, respectively. The measured results, shown in figure~\ref{fig:figure3}, closely follow these predictions, and show a one-to-one relationship between violating the CHSH inequality and observing anomalous (beyond-unit or negative) weak-valued probabilities. The violation and anomalous values occur only for states with high concurrence, above $(1/\sqrt{2}) - 1 \approx 0.41$ (tangle above $3-2\sqrt{2}\approx 0.17$). (Although any pure state with nonzero concurrence can violate the CHSH inequality~\cite{Munro2001}, this requires optimization of the measurement bases, which we do not do here.)

\section{Conclusion}

For our apparatus, the results that we measure can be reproduced by a local hidden variables theory; this is the case with any protocol in which there is no measurement choice on one side~\cite{Saunders12}. We wish to make it very
clear that our experiment is not a Bell test using weak measurements,
nor is it aimed at testing quantum mechanics. However, assuming quantum mechanics to be correct, the negative weak-valued probabilities we measure 
imply that the state could generate strong-measurement correlations which could not be reproduced by any LHV theory.
It is not a new observation that Bell-violating CHSH tests may be explained by recourse to negative joint probabilities~\cite{Feynman91,Spekkens08}. 
What is new here is that such probabilities are not merely hypothetical constructs, chosen arbitrarily from an infinite set of equally valid possibilities, but rather they naturally arise, as weak-valued joint probabilities, derived from the outcomes of experiments probing the intermediate state of the system. Our results open the way to further experimental 
investigations of counter-intuitive quantum phenomena, using weak measurements in place of strong measurements, 
offering more deep insights into the foundations 
of quantum mechanics.

\section*{Acknowledgments}

We thank Holger Hofmann for helpful discussions. This work was supported by the Australian Research Council grant CE110001027.

\end{document}